\documentclass[10pt,reqno,cmintegrals,cmbraces]{amsproc}
\usepackage{titlesec}
\UseRawInputEncoding
\usepackage{newtxtext}
\usepackage{bbold}
\usepackage{anyfontsize}
\usepackage[cmintegrals,cmbraces]{newtxmath}
\titleformat{\section}
{\normalfont\Large\bfseries}{\thesection}{1em}{}
\titleformat{\subsection}
{\normalfont\large\bfseries}{\thesubsection}{1em}{}
\titleformat{\subsubsection}
{\normalfont\normalsize\bfseries}{\thesubsubsection}{1em}{}
\titleformat{\paragraph}[runin]
{\normalfont\normalsize\bfseries}{\theparagraph}{1em}{}
\titleformat{\subparagraph}[runin]
{\normalfont\normalsize\bfseries}{\thesubparagraph}{1em}{}
\usepackage{booktabs} % to put two tables side by side
\usepackage[colorinlistoftodos]{todonotes}
\usepackage{cite}
\usepackage{color}
\definecolor{RED}{HTML}{FF0000}
\usepackage{graphicx}
\usepackage{psfrag,epsfig}
\usepackage{mathrsfs}
\usepackage{tabularx}
\usepackage{xr}
\usepackage{arydshln}
\usepackage{siunitx,array,multirow}
\usepackage{rotating}
\usepackage[letterpaper,margin=1in]{geometry}
\usepackage{subfigure}
\usepackage{enumerate}
\usepackage{comment}
\usepackage{lineno}
\usepackage{algorithmic}
\usepackage{algorithm}
\usepackage[small]{caption}
\usepackage{bbold}
\usepackage[normalem]{ulem}
\usepackage{booktabs}
\usepackage{zref-user}
\usepackage[debug=false, colorlinks=true, pdfstartview=FitV, linkcolor=blue, citecolor=blue, urlcolor=blue]{hyperref}
\hypersetup{
    colorlinks=true,
    linkcolor=blue,
    filecolor=magenta,      
    urlcolor=blue,
}
\usepackage{float}
\mathchardef\mhyphen="2D
\usepackage{chemarr}
\usepackage[version=3]{mhchem}
\newcommand*\patchAmsMathEnvironmentForLineno[1]{%
  \expandafter\let\csname old#1\expandafter\endcsname\csname #1\endcsname
  \expandafter\let\csname oldend#1\expandafter\endcsname\csname end#1\endcsname
  \renewenvironment{#1}%
     {\linenomath\csname old#1\endcsname}%
     {\csname oldend#1\endcsname\endlinenomath}}% 
\newcommand*\patchBothAmsMathEnvironmentsForLineno[1]{%
  \patchAmsMathEnvironmentForLineno{#1}%
  \patchAmsMathEnvironmentForLineno{#1*}}%
\AtBeginDocument{%
\patchBothAmsMathEnvironmentsForLineno{equation}%
\patchBothAmsMathEnvironmentsForLineno{align}%
\patchBothAmsMathEnvironmentsForLineno{flalign}%
\patchBothAmsMathEnvironmentsForLineno{alignat}%
\patchBothAmsMathEnvironmentsForLineno{gather}%
\patchBothAmsMathEnvironmentsForLineno{multline}%
}

\title{Quantifying local and global mass balance errors in physics-informed neural networks}
\author[Mamud et al.,]{M.~L.~Mamud$^{1^{\dagger},*}$, M.~K.~Mudunuru$^1$, S.~Karra$^2$, and B.~Ahmmed$^{3}$\\
{\small $^1$Subsurface Science Group, Pacific Northwest National Laboratory, Richland, WA 99352.\\ $^*$Corresponding author:~Md Lal Mamud, Email:~\texttt{lal.mamud@pnnl.gov}} \\
{\small $^{{\dagger}}$ Previously (GRA) at Earth and Environmental Sciences Division, Los Alamos National Laboratory, Los Alamos, NM 87545.} \\
{\small $^{{\dagger}}$ Previously (GRA) at Geology and Geological Engineering, University of Mississippi, University, MS 38677.}\\
{\small $^2$Environmental Molecular Sciences Laboratory, Pacific Northwest National Laboratory, Richland, WA 99352.} \\
{\small $^3$ Earth and Environmental Sciences Division, Los Alamos National Laboratory, Los Alamos, NM 87545.}}

\date{\today}

\begin{document}
\maketitle

%\linenumbers
\raggedbottom

\textbf{Abstract:}~Physics-informed neural networks (PINN) have recently become attractive for solving partial differential equations (PDEs) that describe physics laws. 
By including PDE-based loss functions, physics laws such as mass balance are enforced softly in PINN. 
This paper investigates how mass balance constraints are satisfied when PINN is used to solve the resulting PDEs.
We investigate PINN's ability to solve the 1D saturated groundwater flow equations for homogeneous and heterogeneous media and evaluate the local and global mass balance errors.
We compare the obtained PINN's solution and associated mass balance errors against a two-point finite volume numerical method and the corresponding analytical solution. 
We also evaluate the accuracy of PINN in solving the 1D saturated groundwater flow equation with and without incorporating hydraulic heads as training data.
We demonstrate that PINN's local and global mass balance errors are significant compared to the finite volume approach.
Tuning the PINN's hyperparameters, such as the number of collocation points, training data, hidden layers, nodes, epochs, and learning rate, did not improve the solution accuracy or the mass balance errors compared to the finite volume solution. 
Mass balance errors could considerably challenge the utility of PINN in applications where ensuring compliance with physical and mathematical properties is crucial.
\section{Introduction}
\label{Sec: introduction}
Partial differential equations (PDEs) are mathematical expressions that describe physical laws governing natural and engineered systems, such as heat transfer, fluid flow, and wave propagation. 
They are widely used to predict physical variables of interest in spatial and temporal domains. 
The solutions of PDEs are often combined with laboratory and field experiments to help stakeholders make informed decisions. Therefore, accurate estimation of variables of interest is critical. 
Numerous methods are used to solve PDEs, including 
analytical \cite{Guo1997AnalyticalWell, Meleshko2005MethodsEngineering, Wexler1989AnalyticalFlow}, 
finite difference \cite{Richardson1910OnDam ,Peiro2005FiniteEquations}, 
finite volume \cite{LeVeque2002FiniteProblems, Moukalled2016TheMatlab, Peiro2005FiniteEquations}, 
finite element \cite{Peiro2005FiniteEquations, Johnson2009NumericalMethod}.

In recent times, there have been significant developments in machine learning (ML) and physics-informed machine learning methods (PIML), leading to the emergence of physics-informed neural networks (PINN) \cite{Lagaris1997ArtifialEquations, Lagaris2000Neural-networkBoundaries, Raissi2017PhysicsEquations, Chen2022FlowDNN:Prediction, Lu2019DeepXDE:Equations} as a popular tool for solving partial differential equations (PDEs).
The approach involves treating the primary variable in the PDE as a neural network. 
The underlying PDE is then discretized using auto-differentiation, and modern optimization frameworks are used to minimize a loss function. The loss function captures losses from the residual PDE and the initial and boundary conditions. 
Such PINN methods have been used to solve a wide range of scientific and engineering problems, including metamaterials, fluid dynamics, and geometries \cite{Raissi2017PhysicsEquations, Lagaris1997ArtifialEquations, Lagaris2000Neural-networkBoundaries, Chen2019Physics-informedMetamaterials,Chen2022FlowDNN:Prediction, Lu2019DeepXDE:Equations, Han2017SolvingLearning, Kakkar2022Physics-InformedGeometries, Liu2019SolvingNetwork, Rezaei2022AMethod, Eivazi2022Physics-informedEquations, Kashefi2022PredictionPointNet, Raissi2017PhysicsEquations, Giampaolo2022Physics-informedSystems}.

These literature examples demonstrate that PINN can handle complex geometry quite effectively. 
However, it has been shown that training a PINN model is a computationally expensive process, and the resulting solutions are not always as accurate compared to numerical solutions \cite{Raissi2017PhysicsEquations, Jagtap2020ConservativeProblems, Chuang2022ExperienceFrustration}. 
For instance, the training process can take thousands of epochs, and the absolute error of forward solutions is approximately $10^{-5}$ due to the difficulties involved in solving high-dimensional non-convex optimization problems \cite{Chuang2022ExperienceFrustration, Cuomo2022ScientificNext}.

Several studies have examined the limitations of PINN in terms of computation time and accuracy. 
For example, Jagtap et al. \cite{Jagtap2020ConservativeProblems} addressed the accuracy issue of PINN by introducing an additional constraint term to the loss function that enforces continuity/flux across the subdomain boundaries. 
However, this additional term makes the model more complex to train, requiring more data and increased computational cost. 
On the other hand, Huang et al. \cite{Huang2023EfficientEncoding} focused on improving the training efficiency of PINN by incorporating multi-resolution hash encoding that provides locally-aware coordinate inputs to the neural network. 
However, this encoding method requires careful selection of hyperparameters and auto-differentiation, complicating training and post-training processes. A comprehensive study is required to evaluate how well PINN solutions satisfy balance laws, such as mass, momentum, and energy. 

% In this study, we investigate the ability of PINN solutions to meet the mass balance requirements. 
% We have selected the application of groundwater flow, where the mass flux is linearly dependent on head gradients through Darcy's model. 

% We employed PINN to solve simple boundary value PDEs describing 1D steady-state saturated groundwater flow in homogeneous and heterogeneous confined aquifers. 
% For comparison, we solved the same governing equations using analytical methods and a traditional numerical method known as the two-point finite volume (FV) method \cite{Moukalled2016TheMatlab}.

This study aimed to determine whether PINN conserved local and global mass and compared its performance with the FV method. 
For this purpose, we solved a steady-state 1D groundwater flow equation to calculate hydraulic heads using analytical, the two-point finite volume (FV), and PINN methods for homogeneous and heterogeneous media cases. 
We conducted hyperparameter tuning by generating an ensemble of 10,800 unique scenarios for each case using a grid search technique to explore all possible combinations of hyperparameters and find the best PINN model. 
Subsequently, we computed the accuracy of the FV and PINN models using performance metrics such as $MSE$, $ R^2$, $MBE$, and $NSE (spatial)$, comparing the hydraulic heads calculated using the analytical solution. 
Finally, we investigated the integrity of the PINN models by computing Darcy's flux, local mass balance error, and global mass balance error.

The paper is structured as follows: Section~\ref{Sec:S2_GE} provides the governing equations for groundwater flow and equations for mass balance error calculation.
Section~\ref{Sec:S3_methods} presents the analytical, two-point flux finite volume, and PINN solutions for the governing equations described in Section~\ref{Sec:S2_GE}, hyperparameters tuning, and performance matrices. 
Section \ref{Sec:S4_results} discusses the results for tuning PINN to find the best model, and compares the analytical, finite volume, and PINN solutions for the hydraulic head, Darcy's flux, and mass balance error. 
Finally, conclusions are drawn in Section \ref{Sec:S5_conclusions}.
%================================;
%  Section: Governing equations  ;
%================================;
\section{Governing Equations}
\label{Sec:S2_GE}

%------------------------;
%  Equation: Continuity  ;
%------------------------;
The one-dimensional steady-state balance of mass for groundwater flow in the absence of sources and sinks within the flow domain is:
\begin{equation} 
    \label{eq:continuity}
     \frac{\partial}{\partial {x}}\left[{{q}(x)}\right] = 0, 
\end{equation}
%--------------------------;
%  Equation: Darcy's flux  ;
%--------------------------;
where, $q(x)$, follows the Darcy's model given by:
\begin{equation} 
\label{eq4:flux_hetero}
    {q}(x) =  - K(x)\frac{{\partial {h}}}{{ \partial {x}}},	
\end{equation}
where, $h[L]$ is the piezometric head,  $x [L]$ is the coordinate, and $K [L/T]$ is the hydraulic conductivity.
%--------------------------------;
%  Equation: governing equation  ;
%--------------------------------;
Using Eqs.~\ref{eq:continuity} and \ref{eq4:flux_hetero}, the governing equation for the balance of mass becomes:
\begin{equation} \label{eq:ge_hetero}
     \frac{\partial}{{\partial {x}}}\left[ {K\left( x \right)\frac \partial {{\partial x}}h(x)} \right] = 0. 
\end{equation}

Eq.~\ref{eq:ge_hetero} requires two boundary conditions for the head. 
Assuming a unit-length domain, if $h_L, h_R$ are the heads at the left and right faces of the domain, then $h(x=0) = h_L$ and $h(x=1) = h_R$.

\begin{figure}[!htbp]
%\begin{figure}[h]
     \centering
        {\includegraphics[width = 0.495\textwidth]
        {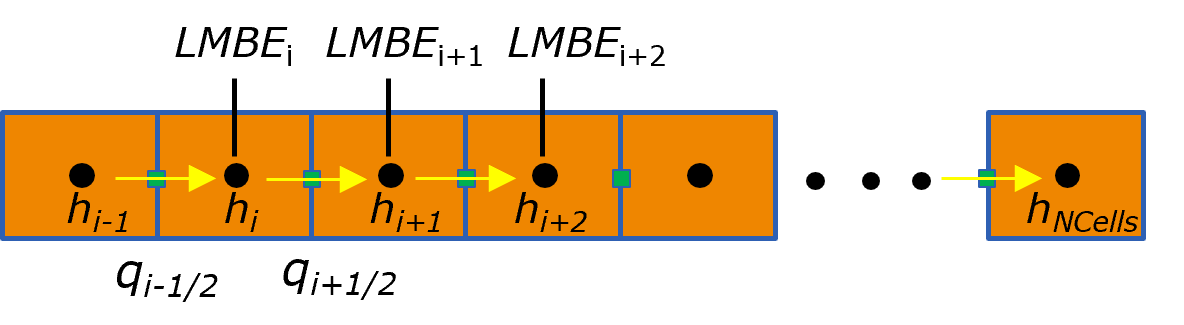}}
        \caption{Conceptual 1D model illustrating the locations of the head, Darcy's flux, and local mass balance error ($LMBE$) calculations.}
   \label{fig:conceptual_model}
 \end{figure}

The model domain ($x \in [0,1]$) is discretized into a finite number of cells ($NCells$).
Figure~\ref{fig:conceptual_model} shows the discretization for obtaining both the FV and PINN solutions.
Assuming a unit area for the one-dimensional domain, the mass flux at face $i-\frac{1}{2}$ ($m_{\mathrm{local}_{i-\frac{1}{2}}}$), the local mass balance error (\textit{LMBE}) in cell $i$, and the global mass balance error (\textit{GMBE}) are: 
\begin{subequations}
  \begin{align}
    \label{eq:lmf}
    m_{\mathrm{local}_{i-\frac{1}{2}}} &= \rho_w q_{i-\frac{1}{2}}, \\
    \label{eq:lmbe}
    LMBE_i &= {m_\mathrm{local}}_{i + \frac{1}{2}} - {m_\mathrm{local}}_{i - \frac{1}{2}} \quad \forall i = 1, 2, \cdots, NCells,\\  
    \label{eq:gmbe}
    GMBE &= \sum \limits_{i = 1}^{NCells} LMBE_i,
  \end{align}
\end{subequations}
where, ${\rho _w}$ $[M/L^3]$ is the density of water, $q_{i-\frac{1}{2}}$ $[L/T]$ is Darcy's flux at face $i-\frac{1}{2}$, $x_{i-1/2}$ and $x_{i+1/2}$ are locations of the faces where the Darcy fluxes are computed.

% ------------------------------------------------------------;
%  Figure-1: Conceptual 1D model for local and global fluxes  ;
% ------------------------------------------------------------;

%========================;
%  Section: Methodology  ;
%========================;
\section{Methodology}
\label{Sec:S3_methods}
This section briefly describes the analytical, FV, and  PINN methods to solve Eq.~\ref{eq:ge_hetero}. 
We consider two scenarios -- homogeneous ($K$ is a constant) and heterogeneous ($K$ varies with $x$) porous media. 

%-----------------------------------;
%  Subsection: Analytical solution  ;
%-----------------------------------;
\subsection{Analytical solution}
\label{subSec:as}
Considering $K$ as constant and choosing $h(x=0) = 0$ and  $h(x=1) = 0.9$ as the Dirichlet boundary conditions, the analytical solution can be derived as follows:
\begin{equation} \label{eq6:as_homo}
    h =  - 0.1x + 1.
\end{equation}
With $K$ varies as a function of space as $ K(x) = (x + 0.5)^2/2.25$ (a non-negative convex function), the analytical solution for the heterogeneous scenario can be derived as:
\begin{equation} \label{eq7:as_hetero}
    h = \frac{{0.075}}{{\left( {x + 0.5} \right)}} + {0.85}.
\end{equation}

%----------------------------------;
%  Subsection: Numerical solution  ;
%----------------------------------;
\subsection{Numerical solution using finite volume}
\label{subsec:FDM}
For the homogeneous medium, the discretized forms of Eq.~\ref{eq:ge_hetero} using the two-point flux finite volume with a central difference for gradient, $\frac{dh}{dx}$, are as follows: 
\begin{equation}
\label{eq8:ns_homo}
    {h_{i + 1}} - 2{h_i} + {h_{i - 1}} = 0,
\end{equation}
while for the heterogeneous medium: 
\begin{equation}
\label{eq9:ns_hetero}
    \left( {{K_{i + 1}} + {K_i}} \right){h_{i + 1}} - \left( {{K_{i + 1}} + 2{K_i} + {K_{i - 1}}} \right){h_i} + \left( {{K_i} + {K_{i - 1}}} \right){h_{i - 1}} = 0.
\end{equation}
Along with the boundary conditions, $h(x=0) = 0$, and  $h(x=1) = 0.9$, the linear systems of Eqs.~\ref{eq8:ns_homo} and \ref{eq9:ns_hetero} were solved using the LU decomposition method in \textsf{Scipy} python package \cite{Virtanen2020SciPyPython}.

%------------------------------;
%  Subsection: PINN solution  ;
%------------------------------;
\subsection{PINN solution}
\label{Sec:PINN_solutions}
The deep neural networks (DNN) approximation of groundwater heads as a function of $x$, and weights and biases ($\theta$) of a neural network is: 
\begin{equation}\label{eq11:pinn_head}
    h(x) \approx \hat h(x;\theta ).
\end{equation}
DNN approximation of the governing Eq.~\ref{eq:ge_hetero} for the homogeneous case is given by: 
\begin{equation}\label{eq12:pinn_dnn_homo}
 	f(x) = \frac{\partial^2{h}} {\partial {x^2}} \approx \hat f(x;\theta ) = \frac{{{\partial^2}\hat h(x;\theta )}}{{\partial{x^2}}},
\end{equation}
while the heterogeneous case is
\begin{equation}\label{eq13:pinn_dnn_hetero}
 	f(x) = \frac{\partial}{{\partial {x}}}\left[ {K\left( x \right)\frac \partial {{\partial x}}h(x)} \right] \approx \hat f(x;\theta ) = \frac{\partial}{{\partial {x}}}\left[ {K\left( x \right)\frac \partial {{\partial x}} \hat h(x;\theta )} \right].
\end{equation}

The loss function (accounting for the losses in the PDE residual and the residual due to the boundary conditions) without training data for the above scenarios is 
\begin{equation}
    \label{eq14:pinn_no_data_homo}
    \mathcal{L}(\theta ) = \frac{1}{{{N_c}}}{\sum\limits_{i = 1}^{{N_c}} {\left[ {\hat f(x_i^c;\theta )} \right]} ^2} + \frac{1}{{{N_D}}}{\sum\limits_{i = 1}^{{N_D}} {\left[ {\hat h\left( {x_i^D;\theta } \right) - g_i^*} \right]} ^2}.
\end{equation}

The first term on the right-hand side in Eq.~\ref{eq14:pinn_no_data_homo} is the loss due to the PDE residual at collocation points. 
The second term in Eq.~\ref{eq14:pinn_no_data_homo} is loss due to the Dirichlet boundary condition. 
In this loss function, ${N_c}$ and ${N_D}$ represent the number of collocation points and the number of Dirichlet boundary conditions, respectively. $x_i^C$ and $x_i^D$ are the locations of the collocation points and Dirichlet boundaries, respectively.

If we include training data for the heads, an additional loss term is added in Eq. ~\ref{eq14:pinn_no_data_homo}, and the overall loss function for this scenario become
\begin{equation}
    \label{eq15:pinn_data_homo}
    \mathcal{L}(\theta ) = \frac{1}{{{N_c}}}{\sum\limits_{i = 1}^{{N_c}} {\left[ {\hat f(x_i^c;\theta )} \right]} ^2} + \frac{1}{{{N_D}}}{\sum\limits_{i = 1}^{{N_D}} {\left[ {\hat h\left( {x_i^D;\theta } \right) - g_i^*} \right]} ^2}{\rm{ + }}\frac{1}{{{N_h}}}{\sum\limits_{i = 1}^{{N_h}} {\left[ {\hat h\left( {x_i^h;\theta } \right) - h_i^*} \right]} ^2},
\end{equation}
This last term in Eq.~\ref{eq15:pinn_data_homo} is the loss to match $h(x)$ to the measurements $h^{*}$.
${N_h}$ represents the number of head measurements/data.  
The head training data and Dirichlet boundary conditions are represented by $h^*$ and $g^*$. 
$x_i^C, x_i^D, x_i^h$ are the locations of the collocation points, Dirichlet boundaries, and head data, respectively.

\begin{figure}[!htbp]
 \centering
{\includegraphics[width = 0.995\textwidth]
{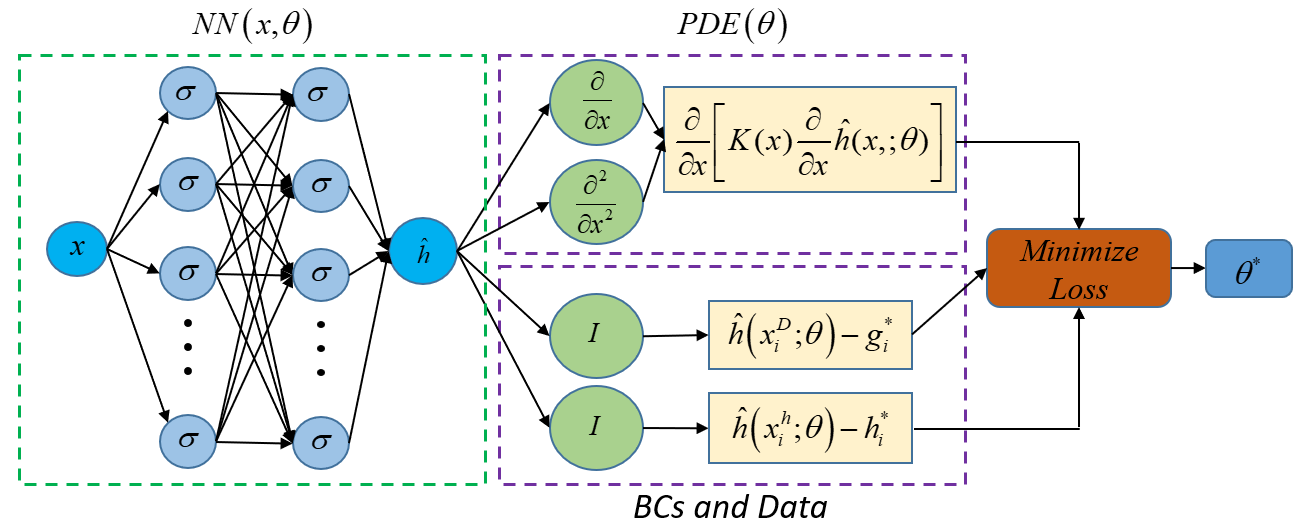}}    
\caption{PINN structure for solving partial differential equation describing 1D steady-state groundwater flow problems in homogeneous and heterogeneous porous media.}
\label{fig:pinn_architecture}
 \end{figure}

Figure \ref{fig:pinn_architecture} illustrates the PINN structure used in this study to solve partial differential equations describing 1D steady-state groundwater flow problems in homogeneous and heterogeneous porous media. The PINN solution, $\hat h(x;\theta)$ (see Figure~\ref{fig:pinn_architecture}), is obtained by
%-----------------------------------------------------;
%  Equation: Minimization problem for PINN training  ;
%-----------------------------------------------------;
\begin{align}
  \label{Eqn:Min_Loss}
  \mathop{\mbox{minimize}}_{\hat h(x;\theta) \in \mathbb{R}^{+}} 
  \mathcal{L}(\theta ),
\end{align}
and the optimal solution is obtained by selecting the model with minimal loss through extensive hyperparameter tuning.

%-----------------------------------------------------;
%  Figure: PINN architecture for solving the problem  ;
%-----------------------------------------------------;

To develop PINN, multi-layer feed-forward DNNs \cite{Bebis1994Feed-forwardNetworks, Svozil1997IntroductionNetworks} are used.
The hyperbolic tangent was used as the activation function because it is infinitely differentiable and facilitated for differentiating Eq.~\ref{eq:ge_hetero} twice.
The remaining PINN hyperparameters are collocation points, learning rate, training data, and epoch numbers. 
Collocation points ($x_i^C$) in Eq.~\ref{eq15:pinn_data_homo} are the locations in the model domain where PINN is trained for satisfying the PDE.
We coincide these points with the FV cell centers for comparison its solution with analytical and numerical solutions.
Training data are the observations/measurements or sampled values from an analytical solution used to train PINN. 
Epoch is one complete pass of the training dataset (PDE, BCs, ICs where needed, and the training data) through the optimization algorithm to train PINN. 
The learning rate is the step size at each epoch while PINN approaches its minimum loss. 
Our codes for PINN used the \texttt{JAX} python package \cite{Bradbury2018JAX:Programs}.

\subsection{Hyperparameter tuning to find the best PINN model}
To evaluate the performance of PINN solutions, we conducted detailed hyperparameter tuning. 
Hyperparameters are settings that control the PINN's learning process, such as the learning rate in an optimization algorithm.
They are distinct from process model parameters (e.g., permeability) learned from the data. 
Our approach involved varying several hyperparameters to generate an ensemble of 10,800 unique scenarios using grid search, which explores all possible combinations of hyperparameters to find the best PINN model within the search space. 
In grid search, we specified a grid of possible values for each hyperparameter you want to tune to define the search space. 
The search algorithm trains the PINN model on all possible combinations of hyperparameter values from the defined grid. The optimal set combines hyperparameter values that lead to the best performance metric (see Sec.~\ref{SubSec:Performance_Metrics}).
We aimed to identify the optimal PINN models for homogeneous and heterogeneous cases.
We varied the hyperparameters, including the number of hidden layers, the number of nodes per hidden layer, the learning rates, the number of epochs, the number of collocation points, and the number of training data points for the head. 
Table~\ref {table: hyperparameters_tuning} lists the values chosen for these parameters.
To run 10,800 realizations, we used a high-performance computing cluster called Tahoma \cite{Laboratory2023Tahoma}, which enabled us to distribute the evaluation of different hyperparameter combinations across multiple processors. 
Tahoma has 184 Intel Cascade Lake nodes, each with a clock speed of 3.1 GHz. 
Each compute node has two 18-core sockets and 36 Intel Xeon Gold 6254 cores per node, and 384GB of memory, giving 10 2/3 GB per core.
By distributing the search, we could reduce the overall tuning time to less than 24 hours to find the optimal hyperparameter set.

%--------------------------------------------------;
% Table for training Parameters for training PINN  ;
%--------------------------------------------------;
\begin{center}
\begin{table}[H]
\caption{The chosen hyperparameters and their ranges for finding the best PINN model for homogeneous and heterogeneous scenarios.
We employ grid-search to train 10,800 different DNN architectures for both scenarios.
}
\label{table: hyperparameters_tuning}
\begin{tabular}{@{}ll@{}}
\toprule
Hyperparameter                     &  Values         \\ \midrule
Number of hidden layers            & $1$, $2$, $3$                                                           \\
Number of nodes                    & $10$, $25$, $50$,                                                       \\
Learning rate                      & $10^{-4}$, $10^{-3}$, $10^{-2}$, $10^{-1}$                              \\
Number of epochs                   & $(10, 20, 30, 40, 50, 60, 70, 80, 90, 100) \times 10^3$                 \\
Number of collocation points       & $11, 21, 41, 81, 161$                                                   \\
Number of training data points     & $0, 7, 20, 40, 80, 160$                \\ \bottomrule
\end{tabular}
\end{table}
\end{center}

\subsection{Performance Metrics}
\label{SubSec:Performance_Metrics}
In addition to the loss function, we used mean squared error ($MSE$), correlation coefficient ($R^2$), mean bias error (MBE), and Nash–Sutcliffe's efficiency (spatial version of NSE) performance metrics \cite{Maroufi2021AEquation, Gueymard2014AProjects, Behar2015APlants, Sanikhani2018AMethods} to evaluate the performance of the PINN and FV methods. 
The $MSE$, $R^2$, $MBE$, and $NSE$ are given by:
\begin{subequations}
  \begin{align}
    \label{eq:mse}
    MSE = \frac{1}{n} \sum\limits_{i = 1}^n (h_{\mathrm{a}_i} - h_{\mathrm{p}_i})^2, \\
    %%
    %\label{eq:R-2_score}
    R^2 = 1 - \frac{\sum\limits_{i = 1}^n {{{\left( {{h_{\mathrm{a}_i}} - h_{\mathrm{p}_i}} \right)}^2}}}{\sum\limits_{i = 1}^n {{{\left( {{h_{\mathrm{a}_i}} - {{\bar h}_\mathrm{a}}} \right)}^2}}},\\  
    %%
    %\label{eq:MBE}
    MBE = \frac{1}{n \cdot \bar{h}_{\mathrm{a}}} \times \sum\limits_{i = 1}^n (h_{\mathrm{p}_i} - h_{\mathrm{a}_i}),\\
    %%
    %\label{eq:SD}
    % SD = \frac{1}{n \cdot \bar{h}_{\mathrm{a}}} \times \sqrt{\sum\limits_{i = 1}^n n \cdot (h_{\mathrm{p}_i} - h_{\mathrm{a}_i})^2 - \sum\limits_{i = 1}^n (h_{\mathrm{p}_i} - h_{\mathrm{a}_i})^2},\\
    %%
    %\label{eq:nse}
    NSE = 1 - \frac{\sum\limits_{i = 1}^n (h_{\mathrm{p}_i} - h_{\mathrm{a}_i})^2}{\sum\limits_{i = 1}^n (h_{\mathrm{a}_i} - \bar{h}_{\mathrm{a}})^2},
  \end{align}
\end{subequations}
where, $n$ is the number of hydraulic heads, $h_\mathrm{a}$ is the head from the analytical solution, $h_\mathrm{p}$ is the predicted head from the PINN model, and ${\bar h}_\mathrm{a}$ is the mean of heads derived from analytical solution. The values of the RMSE, $R^2$, $MBE$, and spatial version of $NSE$ would be, respectively, 0, 1, 0, 1 for a perfect model \cite{Gueymard2014AProjects, Behar2015APlants}. 
If the $MBE$ is positive, it indicates that the model tends to overestimate the true values. 
If the $MBE$ is negative, it suggests that the model tends to underestimate the true values. 
An $MBE$ of zero implies that, on average, the model predictions are unbiased and align closely with the true values.
\section{Results and Discussion}
\label{Sec:S4_results}
This section will present the results of tuning the PINN's hyperparameters to find the best model. 
We will then compare the tuned model's performance with the analytical and FV solutions. 
Finally, we will assess the mass balance errors associated with the FV and PINN solutions for both homogeneous and heterogeneous cases.

%***************************************************************************************
\subsection{Homogeneous Porous Media} 
\label{SubSec:results_Homogeneous} %***************************************************************************************
%
%~~~~~~~~~~~~~~~~~~~~~~~~~~~~~~~~~~~~~~~~~~~~~~~~~~~~~~~~~~~~~~~~~~~~~~~~~~~~~~~~~~~~~
% hyperparameters tuning results           %
%~~~~~~~~~~~~~~~~~~~~~~~~~~~~~~~~~~~~~~~~~~~~~~~~~~~~~~~~~~~~~~~~~~~~~~~~~~~~~~~~~~~~~
\subsubsection{Tuning PINN for homogeneous media} 
\label{ss_sec_homo_tuning}
\begin{figure}[!htbp]
 \centering
        %\label{fig:head_homo}
        {\includegraphics[width = 0.995\textwidth]
        {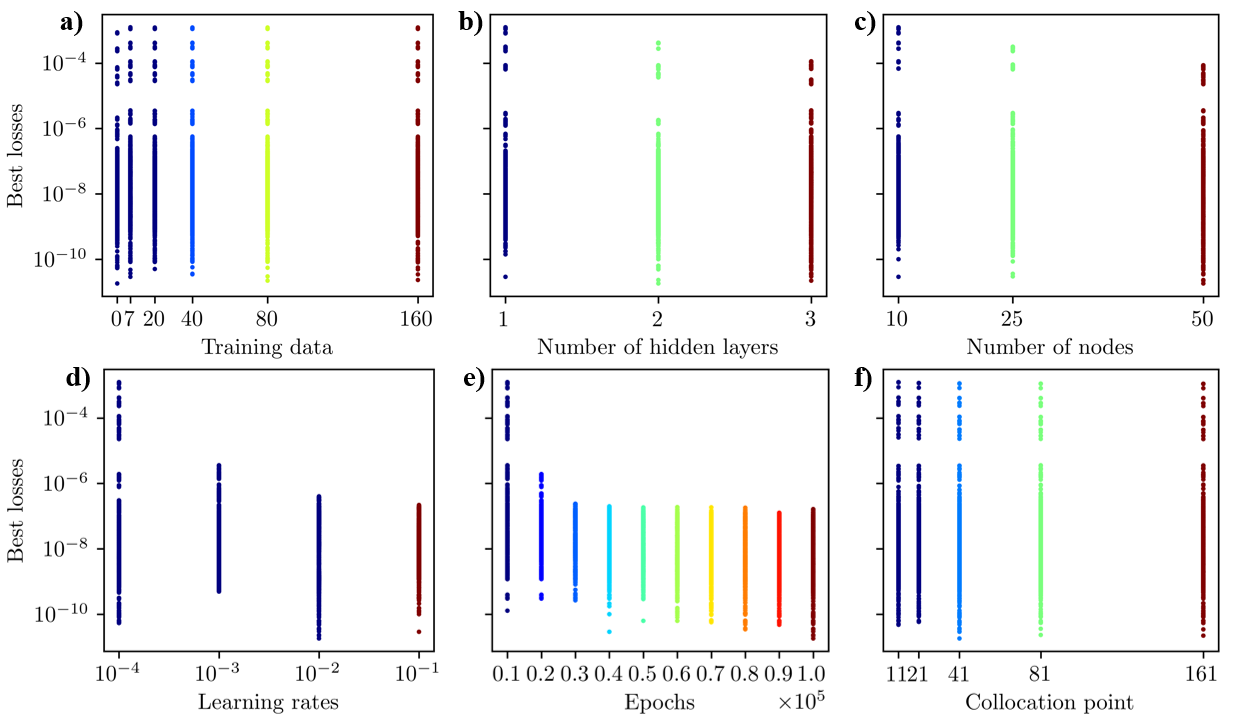}}    
\caption{Best losses for 10,800 hyperparameter scenarios for the homogeneous case using the loss function in Eq.~\ref{eq14:pinn_no_data_homo} and \ref{eq15:pinn_data_homo}. 
The hyperparameters considered were: a) the number of training data points, b) the number of hidden layers, c) the number of nodes per hidden layers, d) the learning rate, e) the number of epochs, and f) the number of collocation points. 
Figures (b), (c), (d), and e show the optimal number of hidden layers, number of nodes per hidden layer, learning rate, and number of epochs are 2, 50, 10$^{-2}$, and 10$^5$, respectively, to obtain the lowest best loss. 
Figure (a) shows that the PINN model with no data outperforms the PINN models with training data to obtain the best loss value.}
\label{fig:homo_finding_best_model}
\end{figure}

We trained physics-informed neural networks (PINN) for 10,800 different hyperparameter scenarios to find the best model for the homogeneous case. 
We used the loss metrics in Eq.~\ref{eq14:pinn_no_data_homo} for PINN without training data and Eq.~\ref{eq15:pinn_data_homo} for PINN with training data. 
The optimal set of hyperparameters was determined by selecting the scenario with the lowest loss value during PINN training.
The hyperparameters considered included the following: the number of training data points (Figure~\ref{fig:homo_finding_best_model}a), the number of hidden layers (Figure~\ref{fig:homo_finding_best_model}b), the number of nodes per hidden layer (Figure~\ref{fig:homo_finding_best_model}c), the learning rate (Figure~\ref{fig:homo_finding_best_model}d), the number of epochs (Figure~\ref{fig:homo_finding_best_model}e), and the number of collocation points (Figure~\ref{fig:homo_finding_best_model}f).
%
% tuning for training data:
In our study, we trained the PINN model using 0, 7, 20, 40, 80, and 160 sets of training data to compare the performance of PINN with and without training data (see Figure~\ref{fig:homo_finding_best_model}a).
The results revealed that the PINN model trained without training data outperformed those trained with datasets. 
A PINN model out of 1,800 trained neural architectures with no data achieved the lowest best loss value of $1.83 \times 10^{-11}$.
The best-performing PINN model trained with data sets out of 9,000 different architectures attained the lowest best loss value of $2.23 \times 10^{-11}$ with 80 training data (see Table \ref {tab:homo_best_pinns_model}). 
Our thorough tuning analysis led us to focus on the best PINN model without training data, providing a robust basis for our discussion.

% tuning for the number of hidden layers:
In Figure~\ref{fig:homo_finding_best_model}b, it is clear that the tuned PINN achieved the lowest losses across scenarios with 1, 2, and 3 hidden layers. 
However, fewer hidden layers resulted in higher values when considering the upper bound of best losses (Figure~\ref{fig:homo_finding_best_model}b). 
More hidden layers led to lower best loss values (Figure~\ref{fig:homo_finding_best_model}b). 
This figure also illustrates that out of 3,600 models, a PINN model with two hidden layers achieved the lowest best loss of $1.83 \times 10^{-11}$ (Table \ref{tab:homo_best_pinns_model}).
%
% tuning for nodes per hidden layers:
We investigated how changing the number of nodes (10, 25, and 50) in each hidden layer affected the optimal loss. 
As shown in Figure~\ref{fig:homo_finding_best_model}c, a PINN model with 50 nodes per hidden layer achieved the lowest loss.
%
% tuning for learning rates:+
Figure \ref{fig:homo_finding_best_model}d shows that using relatively larger learning rates results in significantly lower best losses. 
A low learning rate, such as $10^{-4}$, allows for the exploration of multiple local minima within the non-convex loss function, leading to high values for best loss (Figure \ref{fig:homo_finding_best_model}d). 
The learning rate $10^{-2}$ yielded the best PINN model, achieving the lowest best loss of $1.83 \times 10^{-11}$ (Figure \ref{fig:homo_finding_best_model}d).   
%
% tuning for epoch:
In Figure \ref{fig:homo_finding_best_model}e, it is evident that training the PINN models for a higher number of epochs results in lower best losses. 
For instance, training the PINN for only 10,000 epochs showed many scenarios with higher best losses (Figure \ref{fig:homo_finding_best_model}e). 
The lowest best losses were achieved when the PINN model was trained for 100,000 epochs (Figure \ref{fig:homo_finding_best_model}e).

% -----------------------------------------------------------------------------------
%  Figure: performance analysis and finding the best model for the homogeneous case
% -----------------------------------------------------------------------------------

% -----------------------------------------------------------------------------------------
% Table: Best PINN models with and without data for the homogeneous media 
% ------------------------------------------------------------------------------------------
\begin{table}[!htbp]
\centering
\caption{Hyperparameters and loss of the best two PINN models with and without training data out of 10,800 trained PINN models for the homogeneous case.}
\label{tab:homo_best_pinns_model}
\begin{tabular}{lll}
\toprule
Hyperparameters and Loss     & PINN without training data    & PINN with training data \\
\midrule
Layers              & 2                             & 3 \\
Nodes               & 50                            & 50 \\
Learning Rate       & 0.01                          & 0.01 \\
Best Epoch          & 97,386                        & 91,778 \\
Collocation Points  & 41                            & 161 \\
Training Data       & 0                             & 80 \\
Loss           & $1.83 \times 10^{-11}$        & $2.23 \times 10^{-11}$ \\
\bottomrule
\end{tabular}
\end{table}

\begin{figure}[!htbp]
     \centering
     {\includegraphics[width = 0.995\textwidth]
     {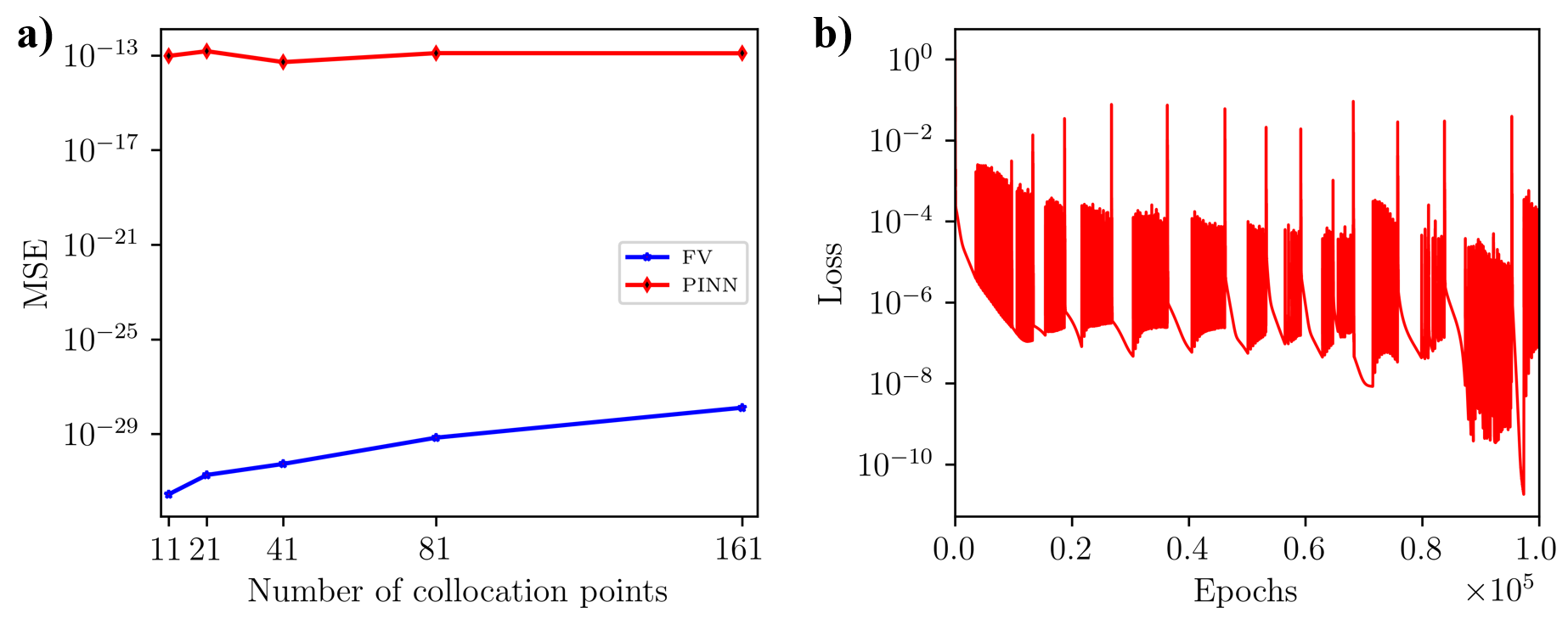}}
   \caption{a) MSE of the FV solutions and best PINN models without training data with the corresponding analytical solutions for all collocation points. 
   Figure (a) shows the optimal collocation points for the PINN model, which is 41, where PINN shows minimum MSE. 
   b) Epoch versus loss values during the best PINN model training. 
   This figure shows that the PINN model achieved the best loss $1.83 \times 10^{-11}$ at 97,386 epoch.}
   \label{fig:homo_loss_convergence}
 \end{figure}

% tuning for collocation points:
Figure \ref{fig:homo_finding_best_model}f illustrates that the number of collocation points has a relatively minor impact on achieving the lowest best loss. Notably, models with 41, 81, and 161 collocation points consistently attained the lowest loss values. 
We investigated the accuracy of the tuned PINN's prediction and the finite volume (FV) solution with the analytical solution to determine the optimal number of collocation points using the mean squared error ($MSE$) metrics.
Figure \ref{fig:homo_loss_convergence}a illustrates that for all the collocation points, the $MSE$ associated with FV solutions converged to double precision ($< 10^{-28}$), while the $MSE$ of PINN predictions with the best PINN models at each collocation point exhibited a relatively higher value of $MSE$ ($\approx 10^{-13}$).
Notably, 41 collocation points achieved the lowest $MSE$ value for the PINN model. 
Based on the above systematic study, we found that the best set of hyperparameters for the PINN model is two hidden layers, 50 nodes per hidden layer, a 0.01 learning rate, 100,000 epochs, 41 collocation points, and no training data.
% -----------------------------------------------------------------------------------
% Figure: FV and PINN convergence with discretization and collocation points
% -----------------------------------------------------------------------------------

% the selected best model training behavior:
Figure \ref{fig:homo_loss_convergence}b shows the evolution of loss over the epochs of the PINN model during training with the best set of hyperparameters. 
This figure indicates that the loss value fluctuates significantly throughout the training process. 
Initially, the loss decreases rapidly, showing that the model is learning quickly and improving its predictions. 
However, periodic spikes in the loss value increase sharply before reducing again. 
This behavior could be due to several factors, including the optimizer's dynamics, learning rate adjustments, or the loss landscape's non-convex nature. 
Despite the frequent fluctuations, the overall long-term trend of the loss is downward, suggesting that the model is gradually improving its performance and reducing the error over time. 
By the end of the training period (at 100,000 epochs), the loss reaches very low values of $1.83 \times 10^{-11}$ at 97,386 epochs. 
This indicates that the PINN has achieved a highly accurate fit to the underlying physical equations.

%~~~~~~~~~~~~~~~~~~~~~~~~~~~~~~~~~~~~~~~~~~~~~~~~~~~~~~~~~~~~~~~~~~~~~~~~~~~~~
% PINN, analytical, and FV solutions          %
%~~~~~~~~~~~~~~~~~~~~~~~~~~~~~~~~~~~~~~~~~~~~~~~~~~~~~~~~~~~~~~~~~~~~~~~~~~~~~
\subsubsection{Hydraulic head solutions for the homogeneous media} 
\begin{figure}[!htbp]
 \centering
        {\includegraphics[width = 0.495\textwidth]
        {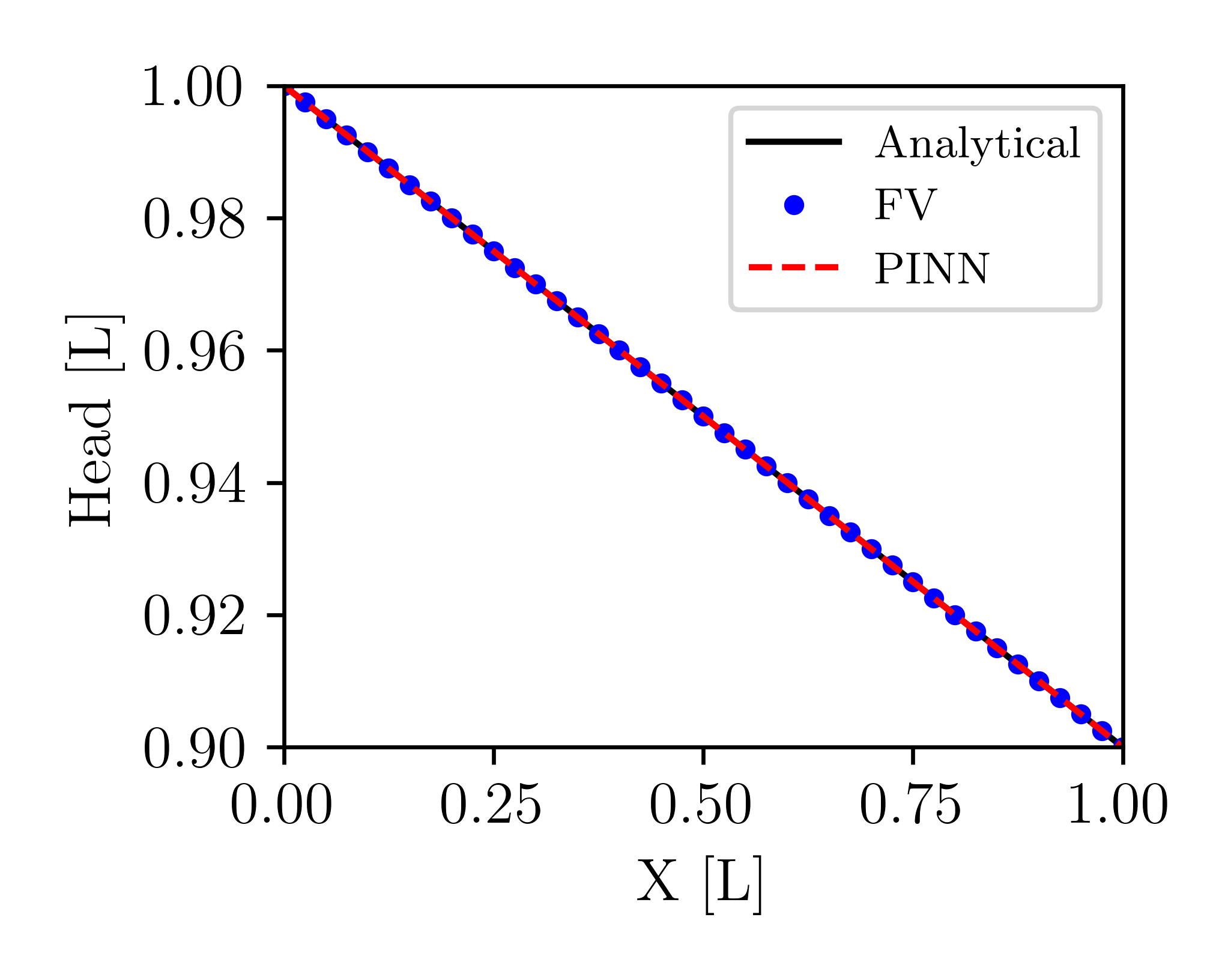}} 
 \caption{Solution of hydraulic heads for the homogeneous porous media of analytical (black line), FV (blue dots), and PINN (red dashes) models.
 %Comparison of the analytical, FV, and best PINN model. 
 %This figure shows that the PINN solution for the 1D steady-state flow problem excellently agrees with the analytical and FV solution
 }
 \label{fig:home_loss_head}
\end{figure}

The hydraulic heads at each collocation point were calculated using the best PINN model with the lowest loss. 
Additionally, the hydraulic heads at these points were computed using analytical and finite volume (FV) methods. 
The FV and the PINN solutions overlap the analytical solution (Figure \ref{fig:home_loss_head}).
Table \ref{tab:homo_FV_pinns_model_performance} presents statistical analysis for performance evaluation of the FV and PINN methods with the corresponding analytical solution.
This table shows that the FV and PINN methods can accurately approximate the true solution for the hydraulic head.
The values of $MSE$, $R^2$, $MBE$, and $NSE (Spatial)$ between the analytical and FV solution are $5.48 \times 10^{-31}$ (approximately the square of machine precision), 1.00, $-1.54 \times 10^{-15}$, and 1.00, respectively. 
These metrics show that the FV solution is identical to the analytical solution, proving its accuracy and reliability. 
The very low $MSE$ and the low $MBE$ confirm that the FV method has minimal errors and biases. 
High $R^2$ and $NSE$ values further confirm the high precision and efficiency of the FV solution in approximating the analytical solution.
Similarly, the values of $MSE$, $R^2$, $MBE$, and $NSE (Spatial)$ between the analytical and PINN predictions are $5.27 \times 10^{-14}$, $1.00$, $-8.92 \times 10^{-7}$, and $1.00$, respectively. 
These metrics collectively indicate that the PINN prediction is accurate and closely matches the analytical solution, similar to the Finite Volume (FV) method. 
Although the $MSE$ for the PINN prediction is slightly higher than that for the FV solution, it is still very low, indicating minimal errors. 
%These results suggest that the PINN model is nearly as accurate and reliable as the FV method.

% -----------------------------------------------------------------------------------
%  Figure: Solution for head and regression of the homogeneous case  ;
% -----------------------------------------------------------------------------------

% ----------------------------------------------------------------------------
% Table: Performance of the FV and best PINN model for the homogeneous media 
% ----------------------------------------------------------------------------
\begin{table}[!htbp]
\centering
\caption{Performance of the FV and PINN with the corresponding analytical solution for the homogeneous case.}
\label{tab:homo_FV_pinns_model_performance}
\begin{tabular}{lllll}
\toprule
Solutions   & $MSE$                     & $R^2$       & $MBE$                         & $NSE (Spatial)$  \\ \midrule
FV          & $5.48 \times 10^{-31}$    & 1           & $-1.54 \times 10^{-15}$       & 1 \\
PINN        & $5.27 \times 10^{-14}$    & 1           & $-8.92 \times 10^{-7}$        & 1 \\ 
\bottomrule
\end{tabular}
\end{table}

% The PINN model with the minimum loss was used to calculate the hydraulic heads at each collocation point.
% Hydraulic heads at the same points as in PINN collocation points were also computed using analytical and FV methods.
% $MSE$ and $R^2$ between the analytical and PINN solution are $4.96 \times 10^{-8}$ and $9.99 \times 10^{-1}$, respectively. 
% Such low MSE and high $R^2$ suggest that the PINN solution is fairly close with the analytical (Figure \ref{fig:home_head_regression}b). $MSE$ and $R^2$ between the analytical and FV solution are $1.68 \times 10^{-32}$ ($\approx$ square of machine precision) and $1.00$, respectively.

\subsubsection{Darcy's flux, local and global mass balance errors}
\begin{figure}[!htbp]
     \centering
     {\includegraphics[width = 0.995\textwidth]
     {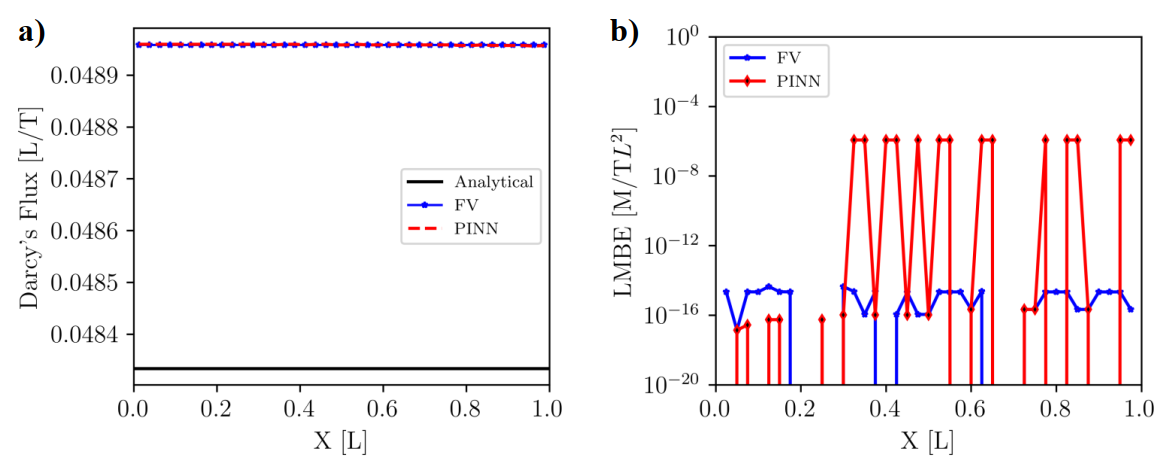}}     
   \caption{a) Darcy's fluxes computed using hydraulic heads with the analytical, FV, and PINN methods for the homogeneous media. Darcy's fluxes of the FV and PINN are higher than that of the analytical solution. 
   b) The LMBE of the FV method is close to machine precision, whereas the LMBE of PINN is higher.}
   \label{fig:homo_flux_mass_balance.PNG}
\end{figure}

Darcy's flux is a key indicator that demonstrates the integrity of a numerical solution in porous media. 
For the steady-state problem, it should remain constant throughout the domain. 
We calculated Darcy's fluxes at all nodes/collocation points within the model domain using Eq.~\ref{eq4:flux_hetero}. 
The Darcy's fluxes computed from the FV and PINN solutions are nearly constant across the solution domain and closer to the analytical solution (Figure~\ref{fig:homo_flux_mass_balance.PNG}a). 
The mean Darcy's flux for the analytical, FV, and PINN methods are $4.833 \times 10^{-2} [L/T]$, $4.896 \times 10^{-2} [L/T]$, and $4.896 \times 10^{-2} [L/T]$ respectively. 
The mean Darcy's fluxes for the FV and PINN methods are slightly higher than that computed from the analytical solution (Table \ref{tab:homo_mass_balance_error}), suggesting numerical errors associated with the FV and PINN methods.
%
% and the magnitude decreases in space.
% However, the mean of Darcy's fluxes computed from the optimal PINN solution was found to be the same as that of the analytical and FV solutions.
%However, Darcy's fluxes of the analytical and FV solutions are constant while they vary for the PINN solution (Figure~\ref{fig:homo_flux_mass_balance}a). 
%They decrease from the beginning to the end of the model domain.   
%The potential reason for such discrepancy has not been explored as it is out of the scope of this study.
%
% -----------------------------------------------------------------------------------
%  Figure-3ab: Darcy's flux and mass balance error plot;
% -----------------------------------------------------------------------------------

The Local Mass Balance Error (\textit{LMBE}) and Global Mass Balance Error (\textit{GMBE}) in the model domain were calculated using \textit{LMBE} and \textit{GMBE} equations for both FV and PINN solutions. 
Analytical solutions are exact solutions that do not incur round-off or numerical truncation errors. 
The \textit{LMBE} and \textit{GMBE} of the analytical solutions are zero due to the exactness of the solutions; therefore, they are not discussed here.
The \textit{LMBE}s of the FV solution range from 0 to $4.46 \times 10^{-15}$ [M/TL$^2$], close to machine precision, with a mean \textit{LMBE} of $1.28 \times 10^{-15}$ [M/TL$^2$]. 
On the other hand, the \textit{LMBE}s of the PINN prediction vary from 0 to $4.46 \times 10^{-6}$ [M/TL$^2$], with an average \textit{LMBE} of $4.18 \times 10^{-7}$ [M/TL$^2]$ (Figure~\ref{fig:homo_flux_mass_balance.PNG}b, Table ~\ref{tab:homo_mass_balance_error}). 
The mean \textit{LMBE} of PINN prediction is eight orders of magnitude higher than that of the FV solution.
The \textit{GMBE} of the FV and the PINN solutions are $5.02 \times 10^{-14}$ and $1.63 \times 10^{-5}$, respectively (Table~\ref{tab:homo_mass_balance_error}). 
The \textit{GMBE} of the PINN predictions is nine orders of magnitude higher than that of the FV solution.

The \textit{LMBE} and \textit{GMBE} for PINN are much higher than those of the FV solution, indicating that PINN do not conserve mass locally and globally (Table 1). 
This discrepancy is primarily due to the fact that FV imposes a strict constraint on the mass balance when solving Eq. \ref{eq:continuity}. 
Although PINN accurately computes heads, it fails to balance the mass because it aims to minimize the residual contributions from the PDE and the BCs instead of setting these residuals to zero. 
Another potential reason for this mismatch is the neural network's inability to optimize its weights and biases as the loss function value decreases due to the non-convex nature of PINN training. 

% -----------------------------------------------------------------------------------------
% Table: Mass balance error with the FV and PINN solution for the homogeneous media 
% ------------------------------------------------------------------------------------------
\begin{table}[!htbp]
\caption{Calculated mean Darcy's flux [L/T], mean and maximum \textit{LMBE} [M/TL$^2$], and \textit{GMBE} [M/TL$^2$] of the analytical, FV, and PINN solutions for homogeneous media.}
\label{tab:homo_mass_balance_error}
\begin{tabular}{lllll}
\toprule
Solutions   & Mean \textit{Darcy's flux}    & Maximum \textit{LMBE}                  & Mean \textit{LMBE}          & \textit{GMBE}\\ 
\midrule
Analytical  & $4.833 \times  10^{-2}$             & 0                                 & 0                          & 0 \\
FV          & $4.896 \times  10^{-2}$             & $4.46 \times  10^{-15}$           & $1.28 \times  10^{-15}$    & $5.02 \times 10^{-14}$\\
PINN        & $4.896 \times 10^{-2}$              & $1.17 \times 10^{-6}$             & $4.18 \times  10^{-7}$    & $1.63 \times 10^{-5}$\\ 
\bottomrule
\end{tabular}
\end{table}

% *************************************************************************************************
\subsection{Heterogeneous Porous Media}
% *************************************************************************************************
%
%~~~~~~~~~~~~~~~~~~~~~~~~~~~~~~~~~~~~~~~~~~~~~~~~~~~~~~~~~~~~~~~~~~~~~~~~~~~~~~~~~~~~~
% hyperparameters tuning results           %
%~~~~~~~~~~~~~~~~~~~~~~~~~~~~~~~~~~~~~~~~~~~~~~~~~~~~~~~~~~~~~~~~~~~~~~~~~~~~~~~~~~~~~
\subsubsection{Tuning PINN for heterogeneous media}
\begin{figure}[!htbp]
 \centering
        %\label{fig:head_homo}
        {\includegraphics[width = 0.995\textwidth]
        {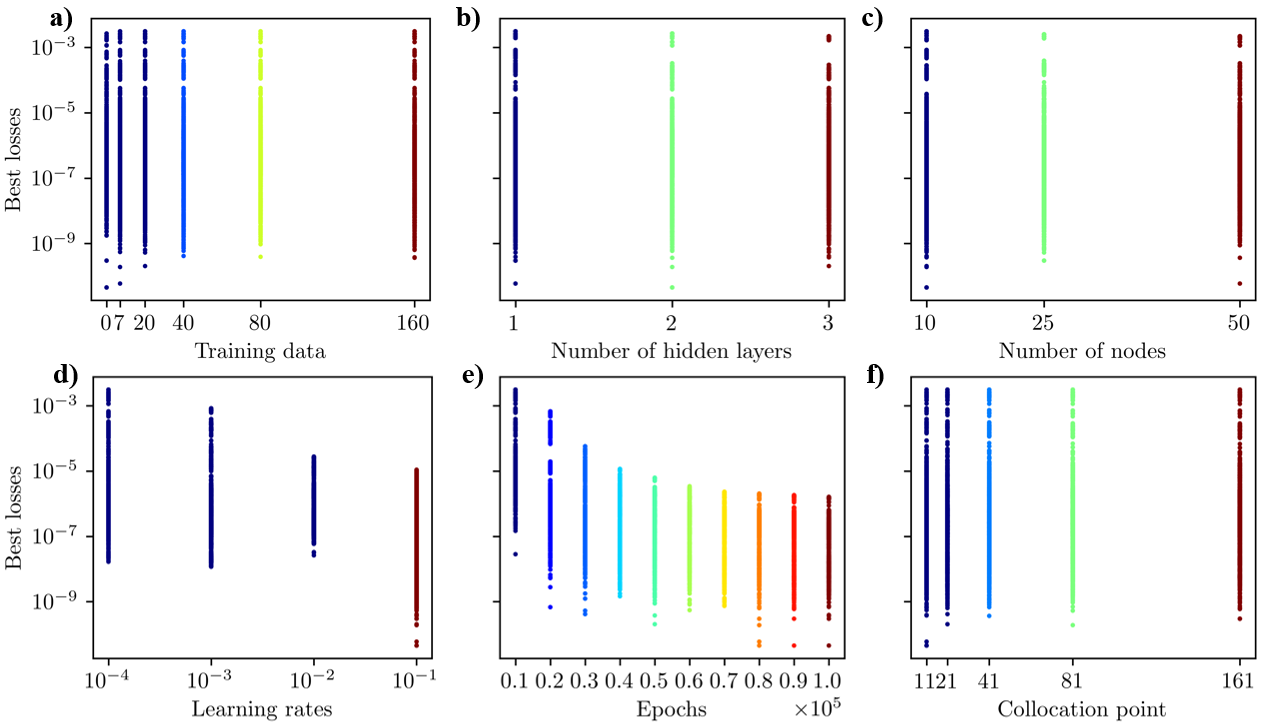}}    
\caption{Best losses for 10,800 hyperparameter scenarios for the heterogeneous case using the loss function in Eq.~\ref{eq14:pinn_no_data_homo} and \ref{eq15:pinn_data_homo}. 
The hyperparameters considered were: a) the number of training data points, b) the number of hidden layers, c) the number of nodes per hidden layer, d) the learning rate, e) the number of epochs, and f) the number of collocation points. 
Figures (b), (c), (d), and (e) show the optimal number of hidden layers, number of nodes per hidden layer, learning rate, and number of epochs are 2, 10, 10$^{-1}$, and 10$^5$, respectively, to obtain the best loss. 
Figure (a) shows that the PINN model with no data performs better than the PINN models with training data to obtain the best loss.}
\label{fig:hetero_finding_best_model}
 \end{figure}

\begin{figure}[!htbp]
    \centering
    {\includegraphics[width = 0.995\textwidth]
     {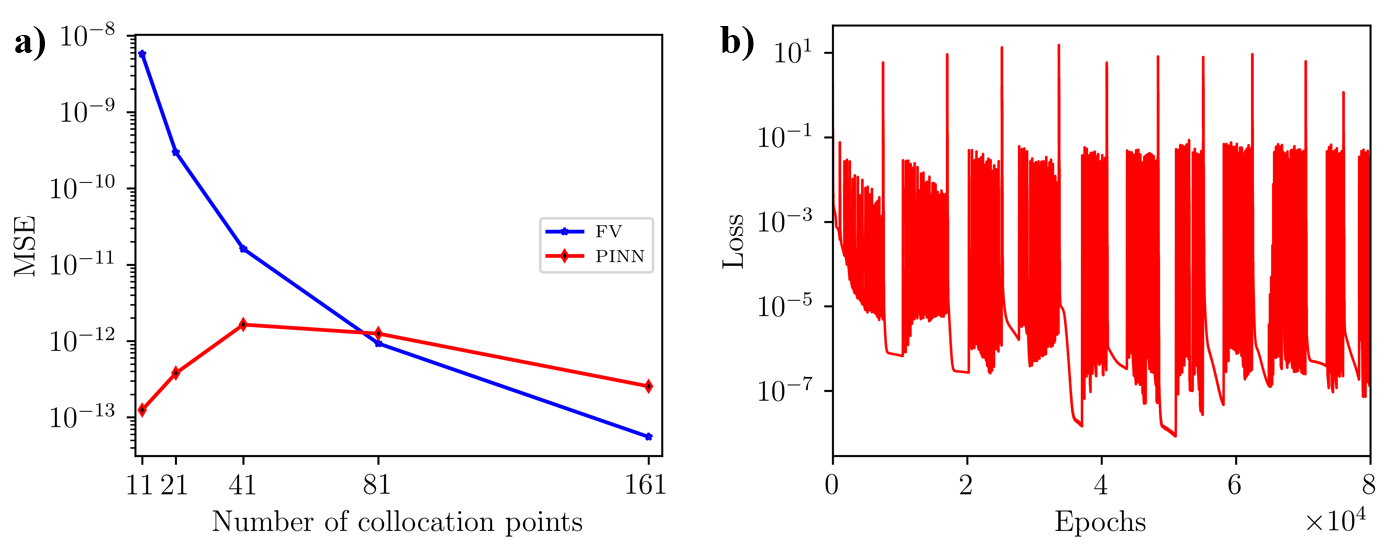}}
    \caption{a) MSE of the FV solutions and best PINN models without training data with the corresponding analytical solutions for all collocation points for heterogeneous media. 
    Figure (a) shows the optimal collocation point for the FV and PINN models 81, where both solutions show comparable MSE. 
    b) Epoch versus loss values during the best PINN model training with 81 collocation points. 
    This figure shows that the PINN model obtained the best loss $4.40 \times 10^{-7}$ at 51,030 epoch.}
    \label{fig:hetero_convergence}
\end{figure}

Similar to homogeneous media, we explored 10,800 hyperparameter scenarios to optimize PINN for the heterogeneous case. 
We utilized the loss metrics outlined in Eq.~\ref{eq14:pinn_no_data_homo} for PINN without training data and the loss function in Eq.~\ref{eq15:pinn_data_homo} for those with training data. 
The optimal hyperparameters were selected based on the lowest loss value during PINN training.
The hyperparameters assessed included the number of training data points (Figure \ref{fig:hetero_finding_best_model}a), the number of hidden layers (Figure \ref{fig:hetero_finding_best_model}b), the number of nodes per hidden layer (Figure \ref{fig:homo_finding_best_model}c), the learning rate (Figure \ref{fig:hetero_finding_best_model}d), the number of epochs (Figure \ref{fig:hetero_finding_best_model}e), and the number of collocation points (Figure \ref{fig:hetero_finding_best_model}f).
%
%Training Data Points:
In our hyperparameter tuning experiments, we trained the PINN model with 0, 7, 20, 40, 80, and 160 sets of training data to evaluate the performance of PINN with and without training data (see Figure~\ref{fig:hetero_finding_best_model}a). 
The results showed that the PINN model without training data performed better than those trained with data sets, similar to the homogeneous case. 
We trained a substantial number of 1,800 models without any data, and one of these models achieved the lowest best loss value of $4.52 \times 10^{-11}$. 
In comparison, the best-performing PINN model with training data sets, out of 9,000 models, reached the lowest best loss value of $5.95 \times 10^{-11}$ with seven training data points (see Table \ref {tab:hetero_best_pinns_model}). 
This comprehensive approach led us to focus on the best PINN model without training data for further analysis.
% Number of Hidden Layers:
In Figure \ref{fig:hetero_finding_best_model}b, it is evident that the best losses for hidden layers 1, 2, and 3 can vary depending on other hyperparameter configurations. 
The data indicates that the PINN models achieved relatively low best losses. 
Specifically, out of 3,600 models with two hidden layers, a PINN model attained the lowest best loss of $4.52 \times 10^{-11}$ (refer to Table \ref{tab:hetero_best_pinns_model} for details). 
%
% Nodes per Hidden Layer:
Adjusting the number of nodes (10, 25, and 50) in each hidden layer has similar effects to changing the number of hidden layers. 
Depending on the other hyperparameter configurations, the best losses may be higher or lower for all the nodes (see Figure~\ref{fig:hetero_finding_best_model}c). 
Out of 3,600 models, a PINN model with ten nodes per hidden layer achieved the lowest best loss (see Figure~\ref{fig:hetero_finding_best_model}c).
%
% Learning Rates:
Figure \ref{fig:hetero_finding_best_model}d shows that using relatively larger learning rates leads to significantly lower best losses. 
For instance, a low learning rate, such as $10^{-4}$, results in higher values for best loss. The best PINN model achieved the lowest best loss of $4.52 \times 10^{-11}$ with a learning rate of $10^{-1}$.
%
% Number of Epochs:
In Figure \ref{fig:hetero_finding_best_model}e, it is evident that training the PINN models for more epochs leads to lower best losses. 
Conversely, training with fewer epochs resulted in significantly higher best losses. 
The PINN models trained for 80,000, 90,000, and 100,000 epochs exhibited similar best-loss values. 
However, the model trained for 80,000 epochs achieved the lowest best losses.
% -----------------------------------------------------------------------------------
%  Figure: performance analysis and finding the best model for the heterogeneous case
% -----------------------------------------------------------------------------------
%
% -----------------------------------------------------------------------------------------
% Table: Best PINN models with and without data for the heterogeneous media 
% ------------------------------------------------------------------------------------------
\begin{table}[!htbp]
\centering
\caption{Optimal values of hyperparameters of the optimal PINN models with and without training data out of 10,800 trained PINN models for the heterogeneous case.}
\label{tab:hetero_best_pinns_model}
\begin{tabular}{lll}
\toprule
Hyperparameters     & PINN without training data    & PINN with training data \\
\midrule
Layers              & 2                             & 1 \\
Nodes               & 10                            & 50 \\
Learning Rate       & 0.1                           & 0.1 \\
Best Epoch          & 76,888                        & 39,399 \\
Collocation Points  & 11                            & 11 \\
Training Data       & 0                             & 7 \\
Loss                & $4.52 \times 10^{-11}$        & $5.95 \times 10^{-11}$ \\
\bottomrule
\end{tabular}
\end{table}
%
%Collocation Points:
The lowest best loss for a PINN model was achieved with 11 collocation points (Figure~\ref{fig:hetero_finding_best_model}f). 
To determine the optimal number of collocation points for comparison, we analyzed the convergence of the PINN prediction and FV solution with the analytical solution. 
As detailed in Eq.\ref{eq:mse}, the mean squared error metric ($MSE$) was used to assess the accuracy.
The $MSE$ of the FV solution decreases as the number of collocation/grid points within the solution domain increases. 
Increasing the number of collocation/grid points allows the FV method to more accurately approximate the continuous partial differential equations governing the physical phenomena, resulting in a decrease in the $MSE$ and an improvement in the solution's accuracy (Figure~\ref{fig:hetero_convergence}a).
Also, the $MSE$ of the best PINN model for 11, 21, and 41 is lower than that of the FV solutions, indicating the spatial resolution independence advantage of the PINN method in predicting accurate solutions (Figure \ref{fig:hetero_convergence}a). 
For further analysis and comparisons of PINN prediction with the FV solution, we selected 81 collocation points where the $MSE$ of both methods is comparable (Figure \ref{fig:hetero_convergence}a). 
The $MSE$ of the PINN prediction and FV for 81 collocation points are $9.27 \times 10^{-13}$ and $1.56 \times 10^{-12}$, respectively. 
Finally, we determined the best optimal hyperparameter values for the PINN model, which includes two hidden layers, 10 nodes, a 0.1 learning rate, 80,000 epochs, 81 collocation points, and 0 training data.
% -----------------------------------------------------------------------------------
% Figure: FV and PINN convergence with discretization and collocation points
% -----------------------------------------------------------------------------------

%
% best model training behavior 
Figure \ref{fig:hetero_convergence}b shows the changes in loss over the epochs of the PINN model during training with the best set of hyperparameters.
The figure demonstrates that the loss fluctuates significantly during training. 
Initially, the loss decreases rapidly, indicating that the model is learning quickly and improving its predictions. 
However, periodic spikes in the loss value occur, increasing sharply before decreasing again. 
At the end of the 80,000-epoch training period, the loss reaches very low values of $8.40 \times 10^{-9}$ at the 51,030-th epoch, indicating that the PINN has achieved a highly accurate fit to the underlying physical equations.

\subsubsection{Hydraulic head solutions for the heterogeneous media} 
\begin{figure}[!htbp]
 \centering
        {\includegraphics[width = 0.495\textwidth]
        {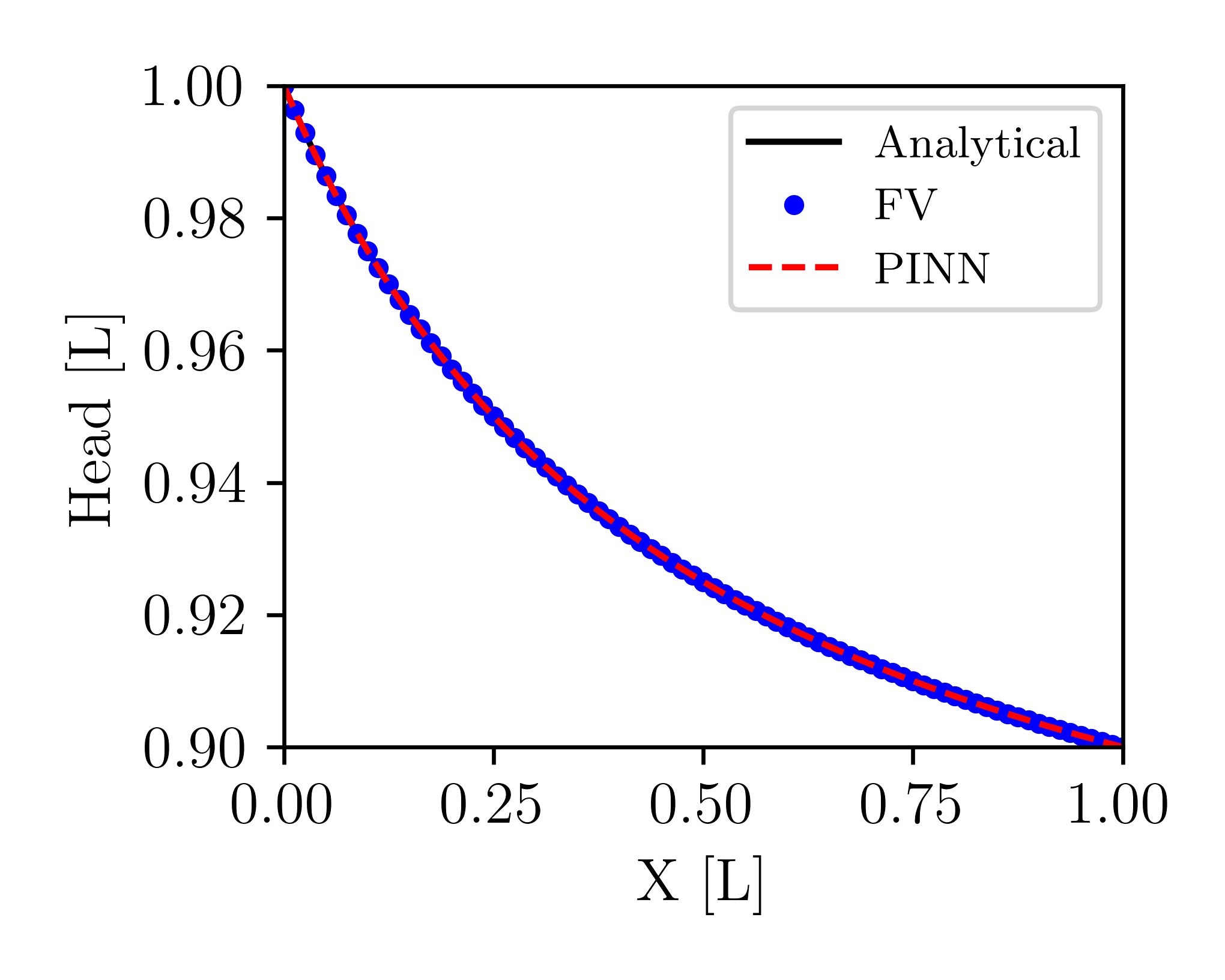}} 
  \caption{Solution of hydraulic heads for the heterogeneous porous media with the analytical, FV, and best PINN model. 
  This figure shows that the PINN solution for the 1D steady-state flow problem excellently agrees with the analytical and FV solution}
 \label{fig:hetero_loss_head}
 \end{figure}

% PINN, analytical, and FV solutions 
%~~~~~~~~~~~~~~~~~~~~~~~~~~~~~~~~~~~~~~~~~~~~~~~~~~~~~~~~~~~~~~~~~~~~~~~~
The best PINN model, which achieved the lowest loss, was used to calculate the hydraulic heads at each collocation point. 
Hydraulic heads at these points were also computed using the analytical and FV methods. 
The FV solution and the PINN prediction closely approximate hydraulic heads as in the analytical solution (Figure~\ref{fig:hetero_loss_head}).
The performance evaluation matrics results reveal that the FV and PINN methods can accurately approximate the true solution for the hydraulic head (Table \ref{tab:hetero_FV_pinns_model_performance}).
For the FV solution, the performance metrics are as follows: the mean squared error ($MSE$) is $9.27 \times 10^{-13}$, the coefficient of determination ($R^2$) is $1.00$, the mean bias error ($MBE$) is $2.09 \times 10^{-6}$, and the Nash-Sutcliffe efficiency ($NSE$) for spatial data is $1.00$. 
These metrics indicate that the FV solution is very close to the analytical solution, demonstrating high accuracy and reliability. 
The low MSE and MBE confirm minimal errors and biases in the FV method. 
The perfect $R^2$ and $NSE$ values underscore the FV solution's precision and efficiency in approximating the analytical solution.
For the PINN prediction, the metrics are: $MSE$ of $1.56 \times 10^{-12}$, $R^2$ of $1.00$, $MBE$ of $3.47 \times 10^{-6}$, and $NSE (\text{Spatial})$ of $1.00$. 
These metrics indicate that the PINN prediction is highly accurate and closely matches the analytical solution. 
Although the MSE for the PINN prediction is slightly higher than that for the FV solution, it remains very low, suggesting minimal errors. 
%The results imply that the PINN model is nearly as accurate and reliable as the FV method.

% Hydraulic head prediction by analytical, FV, and PINN with 40 training data points are consistent (Figure \ref{fig:hetero_head_regression_comparison}). 
% The $MSE$ and $R^2$ between hydraulic head prediction by analytical and optimal PINN with 40 training data points are 0.0 and 1, respectively.
% -----------------------------------------------------------------------------------
%  Figure-5ab: Solution for head and regression of the heterogeneous
% -----------------------------------------------------------------------------------

% ----------------------------------------------------------------------------
% Table: Performance of the FV and best PINN model for the heterogeneous media 
% ----------------------------------------------------------------------------
\begin{table}[!htbp]
\centering
\caption{Statistical analysis for performance evaluation of the FV and PINN with the corresponding analytical solution for the heterogeneous case.}
\label{tab:hetero_FV_pinns_model_performance}
\begin{tabular}{lllll}
\toprule
Solutions   & $MSE$                     & $R^2$       & $MBE$                         & $NSE (Spatial)$  \\ \midrule
FV          & $9.27 \times 10^{-13}$    & 1           & $2.09 \times 10^{-6}$         & 1 \\
PINN        & $1.56 \times 10^{-12}$    & 1           & $3.47 \times 10^{-6}$         & 1 \\ 
\bottomrule
\end{tabular}
\end{table}

%~~~~~~~~~~~~~~~~~~~~~~~~~~~~~~~~~~~~~~~~~~~~~~~~~~~~~~~~~~~~~~~~~~~~~~~~
\subsubsection{Darcy's flux, local and global mass balance error}
\begin{figure}[!htbp]
     \centering
     {\includegraphics[width = 0.995\textwidth]
     {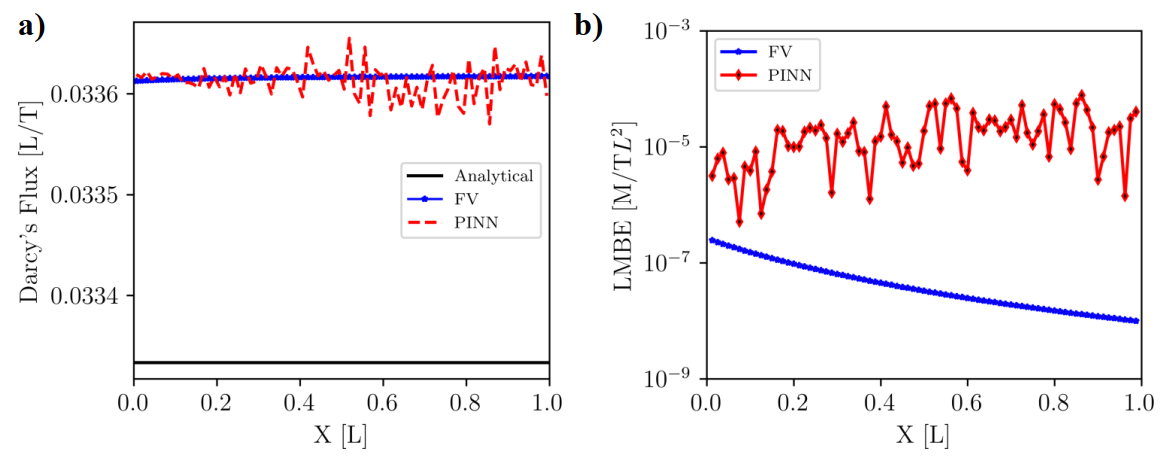}}
      \caption{ a) Darcy's fluxes using hydraulic heads obtained by the analytical, FV, and PINN solutions for the heterogeneous media. 
      Figure (a) shows that Darcy's fluxes of the FV and PINN are higher than that of the analytical solution. 
      Darcy's fluxes of the PINN solution are not constant across the domain. 
      b) Local Mass Balance Error ($LMBE$) of the FV and PINN solutions. 
      This figure shows that the $LMBE$s of the PINN method are higher than that of the FV method}
   \label{fig:hetero_flux_mass_balance}
 \end{figure}

%~~~~~~~~~~~~~~~~~~~~~~~~~~~~~~~~~~~~~~~~~~~~~~~~~~~~~~~~~~~~~~~~~~~~~~~~
We calculated Darcy's fluxes at all nodes/collocation points across the model domain to assess the solution's reliability to the steady-state problem in the heterogeneous case. 
Darcy's flux should remain constant throughout the model domain to uphold the solution in porous media. 
The average Darcy's flux for the analytical, FV, and PINN methods are $3.333 \times 10^{-2} [L/T]$, $3.361 \times 10^{-2} [L/T]$, and $3.361 \times 10^{-2} [L/T]$, respectively (Table \ref{tab:hetero_mass_balance_error}).
Due to numerical error, Darcy's fluxes for the FV and PINN methods are slightly higher than those of the analytical solution (Figure \ref{fig:hetero_flux_mass_balance}).
We observed that while the mean Darcy's fluxes of the FV and PINN solutions are the same, the Darcy's fluxes of the PINN prediction are not constant for the PINN solution (Figure~\ref{fig:hetero_flux_mass_balance}a). 
These results indicate that while PINN can find accurate head predictions for the solution domain, it does not maintain the integrity of the solution.
% -----------------------------------------------------------------------------------
% Figure: Darcy's flux and mass balance error for heterogeneous media
% -----------------------------------------------------------------------------------
% The maximum \textit{LMBE} of the FV and the optimal PINN solutions are $2.44 \times 10^{-7}$ and $3.13 \times 10^{-5}$, respectively (Figure \ref{fig:hetero_flux_mass_balance}b and Table~\ref{tab:hetero_mass_balance_error}). 
% The \textit{LMBE} of the optimal PINN solution is two orders of magnitude higher than that of the FV solution (Figure \ref{fig:hetero_flux_mass_balance}b).
% The \textit{GMBE} of the FV and the optimal PINN solutions are $4.53 \times 10^{-6}$ and $6.92 \times 10^{-6}$, respectively. 
% The magnitudes \textit{GMBE} of PINN and FV are close suggesting the comparable performance of optimal PINN with FV.

The maximum and mean $LMBE$ of the FV solution are $2.43 \times 10^{-7}$ [M/TL$^2$] and $5.73 \times 10^{-8}$ [M/TL$^2$], respectively, whereas the maximum and mean LMBE of the PINN prediction are $7.71 \times 10^{-5}$ [M/TL$^2$] and $2.05 \times 10^{-5}$ [M/TL$^2$] (Figure~\ref{fig:hetero_flux_mass_balance}b, Table ~\ref{tab:hetero_mass_balance_error}). 
The PINN prediction's mean $LMBE$ is three orders higher than the FV solution's.
The \textit{$GMBE$} of the FV and the PINN solutions are $5.02 \times 10^{-14}$ and $1.63 \times 10^{-5}$, respectively (Table ~\ref{tab:hetero_mass_balance_error}). The \textit{GMBE} of the PINN predictions is also three orders of magnitude higher than that of the FV solution. 

The $LMBE$ and $GMBE$ for PINN are higher than those for the FV solution. This suggests that PINN does not conserve mass locally and globally (see Table \ref{tab:hetero_mass_balance_error}).
One of the main reasons for the significant difference in $LMBE$ and $GMBE$ between the FV and PINN solutions is that the FV method strictly enforces a mass balance constraint when solving Eq.~\ref{eq:ge_hetero}. 
Although the PINN model accurately computes head values, it fails to maintain mass balance because its primary goal is to minimize the global residual contributions from the PDEs and BCs rather than setting these residuals exactly to zero.

\begin{table}[!htbp]
\caption{Computed mean $Darcy's flux$ [L/T], mean and maximum \textit{$LMBE$} [M/TL$^2$], \textit{$GMBE$} [M/TL$^2$] of the analytical, FV, and PINN solutions for heterogeneous media.}
\label{tab:hetero_mass_balance_error}
\begin{tabular}{lllll} 
\toprule
Solutions   & Mean \textit{Darcy's flux}          & Maximum \textit{LMBE}             & Mean \textit{LMBE}          & \textit{GMBE}\\ 
\midrule
Analytical  & $3.333 \times  10^{-2}$             & 0                                 & 0                           & 0 \\
FV          & $3.361 \times  10^{-2}$             & $2.43 \times  10^{-7}$            & $5.73 \times  10^{-8}$      & $4.53 \times 10^{-6}$\\
PINN        & $3.361 \times  10^{-2}$             & $7.71 \times 10^{-5}$             & $2.05 \times 10^{-5}$       & $1.58 \times 10^{-3}$\\ 
\bottomrule
\end{tabular}
\end{table}

\section{Conclusions}
\label{Sec:S5_conclusions}
PINN offers an alternative approach to solving physical problems described by partial differential equations. We examined whether PINN preserved local and global mass and assessed its performance against the FV method. This involved solving a steady-state 1D groundwater flow equation to determine hydraulic heads using analytical, FV, and PINN approaches for homogeneous and heterogeneous media scenarios. We fine-tuned hyperparameters by generating 10,800 unique scenarios per case through grid search. Subsequently, we evaluated FV and PINN accuracy using metrics like $MSE$, $ R2$, $MBE$, and $NSE (spatial)$ relative to the analytical solution for hydraulic heads. Lastly, we assessed the integrity of PINN models by analyzing Darcy's flux and local and global mass balance errors.

% homogeneous case
In the case of homogeneous media, we determined the optimal configuration for the PINN model. 
This included two layers with 50 nodes per hidden layer, a learning rate 0.01, 100,000 epochs, 41 collocation points, and no training data through 10,800 hyperparameter tuning scenarios.
The $MSE$, $R^2$, $MBE$, and $NSE (spatial)$ scores were found to be $5.27 \times 10^{-14}$, $1.0$, $-8.92 \times 10^{-7}$, and $1.0$ respectively. 
These scores indicate the reliability and accuracy of the PINN method in predicting hydraulic heads.
Darcy's fluxes with the FV and PINN methods appear similar and relatively constant, resembling the analytical solution. 
However, the fluxes are slightly higher than the analytical solution due to numerical truncation error.
The mean $LMBE$ and $GMBE$ of the PINN prediction are 8 and 9 orders of magnitude higher, respectively, than those of the FV solution. 
These higher values of $LMBE$ and $GMBE$ for the PINN prediction suggest that the PINN method struggles to conserve mass locally and globally while solving a PDE problem.

% heterogeneous case
We have selected the best set of hyperparameters for the PINN model based on convergence analysis and comparison with the FV solution for heterogeneous media case. 
This includes two hidden layers, ten nodes, a learning rate 0.1, 80,000 epochs, 81 collocation points, and no training data for PINN training and prediction.
PINN demonstrated reliability and accuracy in predicting hydraulic head solutions in the heterogeneous case. 
The $MSE$, $R^2$, $MBE$, and spatial $NSE$ scores for the heterogeneous media case are $1.56 \times 10^{-12}$, 1.0, $3.47 \times 10^{-7}$, and 1.0, respectively.
While the mean values of Darcy's fluxes for the finite volume and PINN solutions are consistent, Darcy's fluxes in the PINN prediction fluctuate within the solution domain. 
This fluctuation contradicts the expected behavior, as Darcy's flux should remain constant throughout the solution domain for a steady-state problem. 
The fluctuation of Darcy's flux in the PINN method indicates a lack of integrity in solving the problem.
The mean $LMBE$ and $GMBE$ of the PINN prediction are three orders of magnitude higher than those of the FV solution. 
Similar to the homogeneous media case, PINN fails to conserve mass locally and globally while solving the Partial Differential Equation (PDE) for the heterogeneous media case.
 
% concluding statement
In both homogeneous and heterogeneous cases, the significant difference in $LMBE$ and $GMBE$ between FV and PINN solutions stems from FV's imposition of a strict mass balance constraint, whereas PINN achieves its solution by softly enforcing this constraint when solving PDE problems.
Although PINN calculates accurate heads, it struggles to balance the mass because it aims to minimize the residual contribution from the PDE and the BCs instead of precisely setting these residuals to zero. 
Another potential reason for this discrepancy is the neural network's inability to optimize its weights and biases as the loss function value decreases due to the non-convex nature of PINN training. 
These findings highlight the limitations of PINN in applications where local or global mass conservation is crucial.

%\textit{LMBE}s and \textit{GMBE}s of PINN and FV solutions are around $10^{-6}$. 

%The number of hidden layers and neurons in our neural network architecture was kept constant to investigate the effect of PINN hyperparameters such as epoch, learning rate, collocation point, and training data. 
%Different neural architectures may vary the PINN outputs; however, that will preclude a better understanding of the impact of critical PINN hyperparameters.
%Therefore, keeping neural network architecture is deemed appropriate.
%However, fine-tuning of neural network architecture could be performed to test the PINN performance.
%Based on our extensive PINN simulations, we found that the current loss function is prohibiting learning a better model.
%Whenever the loss function reaches $\approx10^{-5}$ and passes through the back-propagation stage of the learning process, it fails to learn appropriate weights and biases of the neural network.
%We found that PINN mostly learns from boundary conditions and data because they often produce significant errors.
%Alternatively, PDE residuals are often low ($\approx10^{-5}$); therefore, they could not significantly impact the learning process. 
%Therefore, we need a loss function that learns not only from boundary and data but also from PDE residuals efficiently, which can add a significant improvement to PINN.

%=================;
%  ABBREVIATIONS  ;
%=================;
\section*{Abbreviations}
\begin{itemize}
  \item PINN:~Physics-informed Neural Networks
  \item PDE:~Partial Differential Equations
  \item ODE:~Ordinary Differential Equations  
  \item FV:~Finite Volume 
  \item DNN:~Deep Neural Networks 
  \item MSE:~Mean Squared Error
  \item LMF:~Local Mass Flux
  \item LMBE:~Local Mass Balance Error
  \item GMBE:~Global Mass Balance Error 
  \item CP:~Collocation Points
  \item TD:~Training Data
\end{itemize}

%========================;
%  Conflict of interest  ;
%========================;
\section*{Conflict of Interest}
The authors declare that they do not have any conflicts of interest.

%===================;
%  Acknowledgments  ;
%===================;
\section*{Acknowledgments}
MLM and BA thank the U.S. Department of Energy's Biological and Environmental Research Program for support through the SciDAC4 program. 
MLM also thanks the Center for Nonlinear Studies at Los Alamos National Laboratory. 
MLM, SK, and MKM thank the Environmental Molecular Sciences Laboratory (EMSL) for its support. 
EMSL is a DOE Office of Science User Facility sponsored by the Biological and Environmental Research program under Contract No. DE-AC05-76RL01830.
The views and opinions of authors expressed herein do not necessarily state or reflect those of the United States Government or any agency thereof.

\section*{Data \& Codes Availability}
The data and codes used in the paper are available at \url{https://gitlab.com/mlmamud/mass_balance_of_pinn.git} upon publication.
% This repository is currently under Los Alamos National Laboratory's internal review process for public release and will be made available upon approval.
% However, we will make our code available to reviewers upon request.  

\section*{Supplementary File}
The supplementary material contains the derivation of the numerical solutions used in the research.

\bibliographystyle{IEEEtran}
\bibliography{references}

\end{document}